# Performances of a GNSS receiver for space-based applications

Arnaud Dion[1], Vincent Calmettes[1], Michel Bousquet[1], Emmanuel Boutillon[2]
[1]*Université de Toulouse, ISAE, DEOS, TESA, Toulouse, France*
[2]*Lab-STICC, CNRS, Université Européenne de Bretagne, UBS, Lorient France*

## BIOGRAPHY

Arnaud Dion has worked as a development engineer in a small company where he conducted projects based on FPGA. While working, he graduated from CNAM (French engineering school) in 2005. In October 2004, he joined ISAE, the French Aerospace Engineering Institute of Higher Education in Toulouse (formerly SUPAERO), as a project manager in the Department of Electronic, Optronic and Signal (DEOS). His projects include electronic architecture design for a UAV (Unmanned Aerial Vehicle), FPGA and ASIC development, hardware-software codesign. He is also involved in research activities in the frame of a PhD.

Vincent Calmettes received Ph.D. degree in signal processing from SUPAERO (ENSAE), Toulouse, France. He is a training and research scientist at the DEOS. His research interests include the development of solutions based on DSP and programmable logic devices or ASICs for applications in Digital communications and Signal processing. He is currently working on new Galileo signal processing and GNSS receiver architectures. He is also involved in several projects devoted to positioning and attitude determination, including low cost MEMS sensors characterization and INS/GPS integration.

Michel Bousquet is a Professor at ISAE. He is in charge of graduate and post-graduate programmes in aerospace communications and navigation. His research interests cover several aspects of satellite communications and navigation (physical and access layers, system architecture and performance...), within the TeSA co-operative research laboratory (www.tesa.prd.fr). He has been involved in many national and european projects dealing with satellites and he is a member of the Steering Board of SatNEx, the European Network of Excellence in satellite communications.

Emmanuel Boutillon received the Engineering Diploma from the Ecole Nationale Supérieure des Télécommunications (ENST), Paris, France in 1990. In 1991, he worked as an assistant professor in the Ecole Multinationale Supérieure des Télécommunications in Africa (Dakar). In 1992, he joined ENST as a research engineer where he conducted research in the field of VLSI for digital communications. While working as an engineer, he obtained his Ph.D in 1995 from ENST. In 1998, he spent a sabbatical year at the University of Toronto, Ontario, Canada. In 2000, he joined the laboratory Lab-STICC (Université de Bretagne Sud, Lorient, France) as a professor. His current research interests are on the interactions between algorithm and architecture in the field of wireless communications. In particular, he works on Turbo Codes and LDPC decoders.

## ABSTRACT

Space Vehicle (SV) life span depends on its station keeping capability. Station keeping is the ability of the vehicle to maintain position and orientation. Due to external perturbations, the trajectory of the SV derives from the ideal orbit. Actual positioning systems for satellites are mainly based on ground equipment, which means heavy infrastructures. Autonomous positioning and navigation systems using Global Navigation Satellite Systems (GNSS) can then represent a great reduction in platform design and operating costs. Studies have been carried out and the first operational systems, based on GPS receivers, become available. But better availability of service could be obtained considering a receiver able to process GPS and Galileo signals. Indeed Galileo system will be compatible with the current and the modernized GPS system in terms of signals representation and navigation data. The greater availability obtained with such a receiver would allow significant increase of the number of point solutions and performance enhancement. For a mid-term perspective Thales Alenia Space finances a PhD to develop the concept of a reconfigurable receiver able to deal with both the GPS system and the future Galileo system. In this context, the aim of this paper is to assess the performances of a receiver designed for Geosynchronous Earth Orbit (GEO) applications. It is shown that high improvements are obtained with a receiver designed to track both GPS and Galileo satellites. The performance assessments have been used to define the specifications of the future satellite GNSS receiver.

## 1  INTRODUCTION

The SV applications can be divided into 3 categories: Low Earth Orbiter (LEO) which works mainly between 300 and 1000 kilometers above the ground, Geostationary Earth Orbiter (GEO) at around 36000 and kilometers and

Medium Earth Orbiter (MEO) between these two. In this paper, only the GEO applications are considered.

The station keeping is an important phase of the SV life. Due to external perturbations, the trajectory of the SV derives from the ideal orbit. The station-keeping window is the tolerance between the ideal and the real trajectory. This window depends on the performances of the thrusters, the performances of the positioning measurement system and the specifications of the application. Actual measurement systems are based on ground equipments. The typical distance and angular measurement accuracy obtained from these equipments for GEO mission are [1]:
- Semi-major axis (vertical): 60 m.
- Longitude (horizontal): $2.10^{-3}$° i.e. 1470 m.

The operating costs can also be reduced by using autonomous equipment instead of a ground based one.

In a mid-term future, Galileo will be deployed and made available for end-user. Galileo is designed to be inter-operable with GPS and modernized GPS. Both systems will share frequencies and modulations. A receiver that will be able to use the two systems will increase the number of visible satellites. With a better availability of GNSS satellites in the receiving zone, a satellite receiver with classical acquisition and tracking algorithms would be sufficient to compute point position. However, the availability of GPS and Galileo signals will vary on a short range (hours), due to satellite trajectory, and on a long range (years), due to the evolution and modernization of GNSS constellations. The receiver will have to be tolerant to this evolution.

Thales Alenia Space funds a PhD to develop the concept of versatile reconfigurable receiver able to deal with both the GPS system and the future Galileo system. The first phase of the study was to evaluate the feasibility of the concept and to compute the expected performances in order to match the requirements of space applications. A software simulation tool has been used to process the orbital and link budget parameters in order to compute the performances of the radio-frequency (RF) link. In the second phase, a high level model of this receiver has been designed in order to evaluate tracking algorithms and implementation solution.

Hence, this paper presents the performance assessments for a receiver on geosynchronous orbit. Section 2 presents the scenario specifications, based on current spacecraft missions. This section presents also the parameters used to model an inter-satellite communication link between a GNSS satellite and a receiving satellite. The 2 following sections present the parameters that will affect the receiver design: firstly the acquisition threshold and secondly the Doppler span. Finally, section 5 presents the overall performance assessment of the GEO applications.

## 2 MODELLING PARAMETERS

ISAE and the Technical University of Munich have developed a software tool in order to analyze and prepare scientific space missions, such as BayernSat, and also to track satellites, such as HETE2 (launched in 2000). This software, SATORB [2], has been used to process the orbital and link budget parameters in order to compute data such as received C/N0, Doppler, system temperature, atmospheric losses, satellite attitude, etc. These results have then been processed with Matlab to define the specifications of the system and the expected performances. The first task was to define the scenarii specifications and the parameters of the inter-satellite RF link.

### 2.1 Scenarii specifications

The data used in the following study to model the GPS and application satellites trajectory are those of the North American Aerospace Defense Command Two-Line Element (NORAD TLE) sets of 22nd of march 2008. At this date, 30 GPS satellites were active. The Galileo constellation has been modeled with a 27/3/1 walker delta pattern [3]. Therefore, there are 57 GNSS satellites. Many scenarii have been studied such as Telecom2D, Jason, Swift, the International Space Station, etc. However, the application satellite used in this article for performance evaluations is Meteosat9 (see Table 1).

| Parameters | GEO |
|---|---|
| Name | Meteosat9 |
| Orbit | Geostationary |
| Period | 23h56 |
| Semimajor axis (km) | 42166.892 |
| Inclination (deg) | 1.02 |
| RAAN (deg) | 209.006 |
| Spacecraft attitude | Earth pointing |
| Antenna configuration | Nadir pointing, high gain parabolic |

**Table 1: Scenarii specifications**

### 2.2 Link budget parameters

The following analysis uses the transmitting antenna pattern of GPS BlockII presented in [4]. As there is no information available on the Galileo transmitting antenna, the same pattern and the same transmit power as GPS have been considered. The GPS satellites are known to have up to 7 dB margin with respect to the specifications of the transmit power [5]. This will guarantee a security margin for the results obtained in the following study.

In this study, the receiver's antenna is a circular parabolic antenna with a diameter of 0.25 meter (a typical 60% efficiency is assumed). The $\theta_{3dB}$ angle is 56° and the gain in the axis is 9.25 dBi [6].

The noise sources have also to be carefully identified and proportioned. The noise captured by the antenna of the GEO spacecraft is the noise from the earth and from outer space. Under these conditions, the major contribution is that from the earth.

### 2.3 Signals

The receiver is adapted to process the GPS signals as well as the Galileo signals, which are CDMA (Code Division Multiple Access) signals. Other modulations, such as Glonass FDMA (Frequency Division Multiple Access), are not considered in this study. The considered frequencies and signals are listed in Table 2 [7,8,9].

| Constellation | Frequency plan | Frequency (MHz) | Modulation |
|---|---|---|---|
| GPS | L1 | 1575.42 | BPSK-R(1) |
| GPS | L1 | 1575.42 | TMBOC(6,1,4/33) |
| GPS | L2 | 1227.6 | BPSK-R(1) |
| GPS | L5 | 1176.45 | BPSK-R(10) |
| Galileo | E1 | 1575.42 | CBOC(6,1,1/11) |
| Galileo | E5 | 1191.795 | AltBOC(15,10) |

**Table 2: Considered signals**

## 3 ACQUISITION THRESHOLD

The number of visible satellites, and then the ability to compute a point solution, is directly linked to the acquisition and tracking C/N0 thresholds. So the first step is to choose thresholds compatible with the expected performances of the receiver in term of precision of positioning, complexity of implementation, etc.

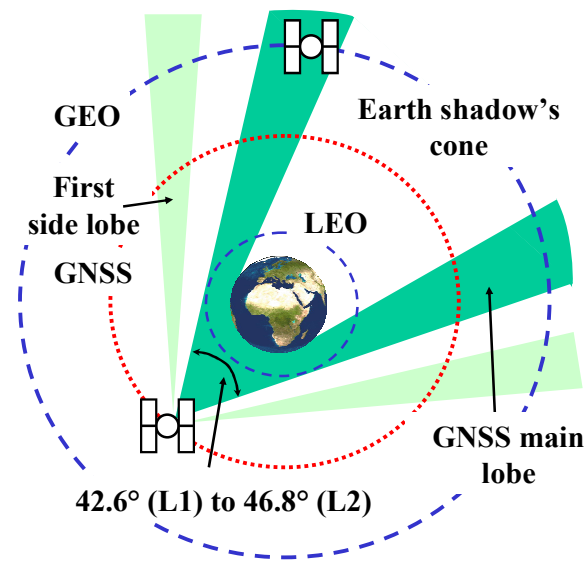

**Figure 1: Geometry for reception of GNSS signals by a GEO spacecraft**

For GEO applications, the most significant constraints are the sparse nature of GNSS signals as well as their low received power. The minimum number of visible GNSS satellites required in order to compute a point solution is four. However, the geostationary satellites are above the GNSS constellations, therefore the receiving zone is dramatically reduced (Figure 1). Typically, four GPS satellites in the visibility domain is a rare event [10].

Figure 2 represents the availability of service, over a 48-hour span, computed for various acquisition thresholds. During these periods, at least 4 satellites must be visible. A satellite is considered here as visible when the received C/N0 is above the acquisition threshold.

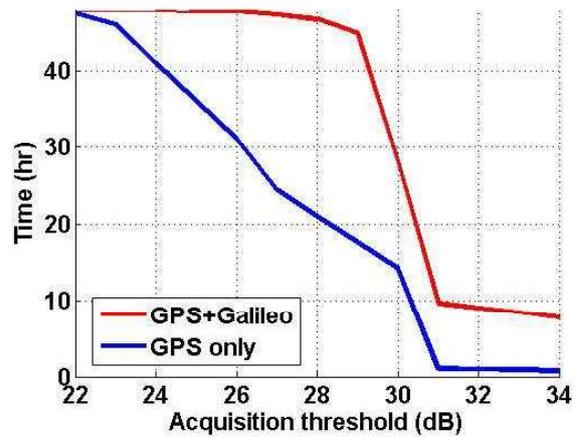

**Figure 2 : Availability of service over 48h for Meteosat9**

This figure is used to specify the acquisition threshold when the 2 constellations are considered. The key value is 29 dBHz. Above this threshold, the total visibility time plummet down but below, the gain is not significant. This figure shows also the benefit of using Galileo along with GPS constellation. This threshold makes sense with the GPS-only configuration, even if the total visibility time is degraded, it is still around 17 hours over 48 hours.

Here, the Geometric Dilution of Precision (GDOP) illustrates the precision performances. It is a purely geometrical factor [3] which doesn't depend on the characteristics of a receiver and gives a good view of the possible performances.

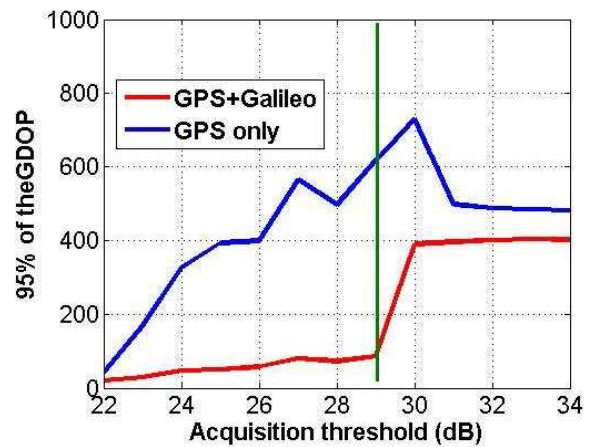

**Figure 3 : 95% of the GDOP for Meteosat9**

Figure 3 presents the 95%-GDOP probable over the acquisition threshold. That means that 95% of the GDOP values, computed over a 48-hour time span, are equal or below that curve at a given threshold. Here again, the benefit of using the Galileo constellation along with the GPS one is clearly showed. At 29 dBHz acquisition threshold, 95% of the GDOP values are below 87. It is compatible with the expected performances for a GEO application. A threshold lower than 29 dBHz will complicate the design of a receiver with little performance improvement. Therefore, this 29 dBHz threshold will be a specification of the future GNSS receiver.

An example of the C/N0 received is presented Figure 4. The acquisition threshold is 29 dBHz and the tracking threshold is 27 dBHz. The side lobes are clearly visible, about 15 dBHz lower than the main lobe, as well as the earth obstruction mask, at the center of the main lobe. GPS signals crossing close to the limb of the Earth are bent or refracted by the atmosphere (ionosphere, troposphere, etc), causing a measurable delay in the signals at the receiver. The most significant error source for a GEO receiver is the delay contributed by the Earth's ionosphere. In our simulations, an atmosphere mask of 1000 km altitude has been used in order to eliminate GPS signals with large atmospheric errors.

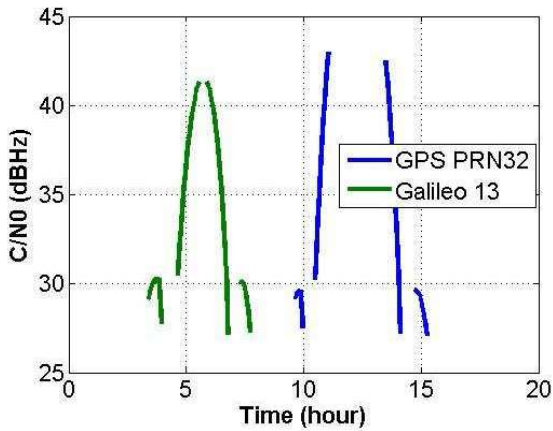

**Figure 4: Meteosat9 C/N0 from GPS PRN32 and Galileo 13 from 22-03-2008 00:00 to 244-03-2008 00:00**

Lots of side lobes signals have a C/N0 ratio below 30 dBHz (see Figure 4). A receiver which is able to exploit the side lobe signals improves the performances. The availability of service is extended (see Figure 2). The GDOP is greatly enhanced (see Figure 3) because the side lobe signals have a greater radial distance from the Earth from the point of view of the receiver (see Figure 5), thus improving the geometry. The skyplot (Figure 5) presents the trajectories of the GNSS satellites, from the point of view of the receiver, when the signals can be tracked, i.e. when the C/N0s are above acquisition and tracking thresholds, respectively, 29 dBHz and 27 dBHz.

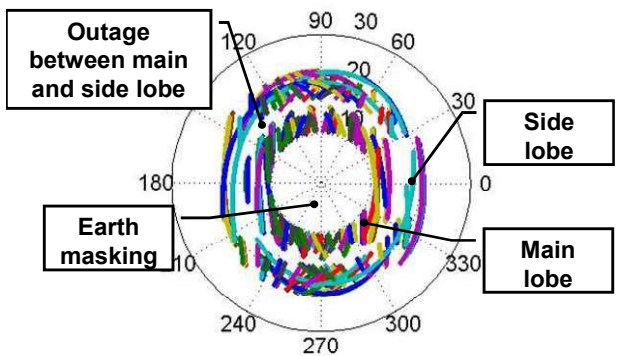

**Figure 5: Skyplot**

## 4 DOPPLER

The Doppler effect of the signals has an impact on the design of the receiver, especially for the acquisition algorithms. For the GEO scenario, the Doppler span is ± 8 KHz (see Figure 6), which is slightly larger than for a terrestrial receiver. But one advantage of space receiver is that the satellite motion is very predictable, hence the search pattern can be considerably reduced in case of a warm start by using the GNSS almanac data.

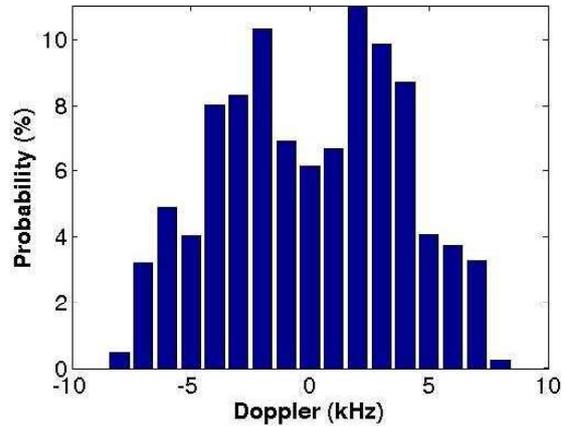

**Figure 6: Distribution of Doppler for Meteosat9**

More than Doppler span, the Doppler rates have an important effect during the tracking phase. The phase and frequency loops have to be robust enough to lock on and to track rapidly changing signals, especially for LEO applications. However, the GEO application presents Doppler rates of $\pm 1$ Hz.s$^{-1}$.

## 5 PERFORMANCES

Adding Galileo constellation to GPS constellation improve the availability of service, as shown Figure 2. It also improves the geometry because there are more side lobe signals that can be demodulated. Figure 7 shows the GDOP improvement over 48h. For GPS only, the mean GDOP is 238 and for GPS and Galileo, it is 98.

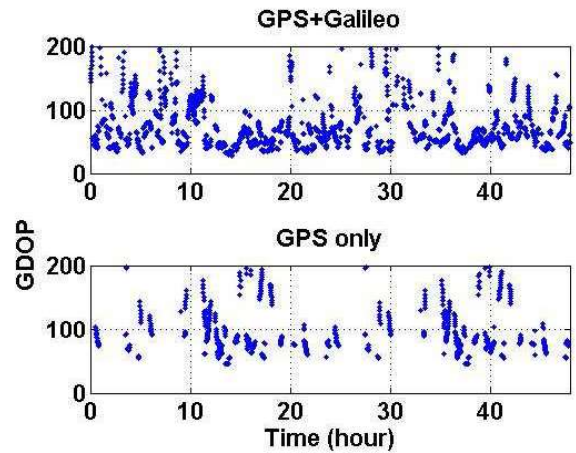

**Figure 7: GDOP over time**

Figure 8 provides a view of the horizontal and vertical standard deviation, calculated over a 48-hour time span.

For GPS only, the vertical error probable (95%) is 398 m and the horizontal error probable (95%) is 67 m for an availability of service of 17h30mn. For GPS and Galileo, these errors are respectively 150m and 25m for an availability of service of 45h. These performances are compatible with the co-localization of several geostationary satellites in the same station-keeping window of ±0.05° [1].

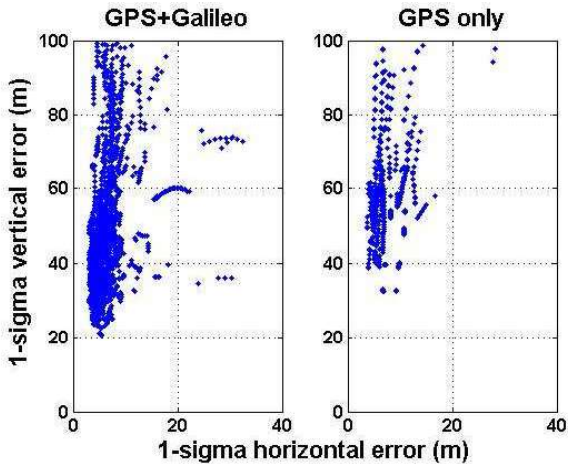

Figure 8: 1-sigma error

For GPS only configuration, the spherical error probable (95%) is 817 m, it means that 95% of the errors in position are below 817 m. For GPS and Galileo configuration, the spherical error probable (95%) is 226 m. Figure 9 shows the position errors assessed with the 2 constellations considered for Meteosat9 over a 48-hour time span.

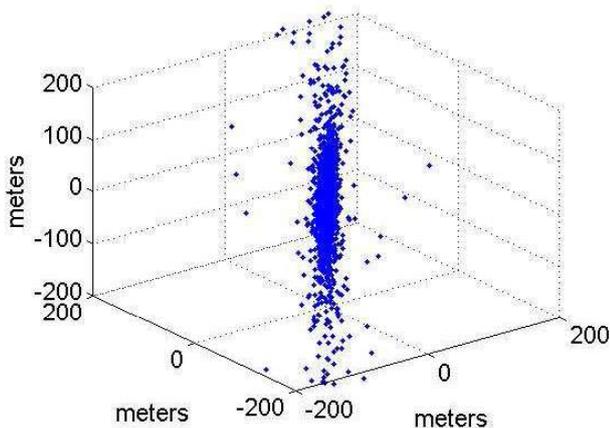

Figure 9: Meteosat9 position error over 48h

This performance is to be compared with typical accuracy presented in the introduction, approximately 1500 meters for measurements with ground-based equipments. Such a performance is compatible with the co-localization of several satellites in the same station-keeping window (ASTRA for example).

## 6  CONCLUSION

We have developed a model which allows the designer to simulate RF links, between GNSS and GEO satellites, and to compute their parameters, such as the C/N0 ratio, the Doppler effect, the ionospheric losses, etc. This approach has been used to define the specifications of a satellite GNSS receiver and to assess its performances. These performances show real improvements compared with current measurement systems. Such a receiver will then be able to provide accuracy compatible with GEO mission.

The main parameters that will affect the design of a satellite receiver are the acquisition threshold and the Doppler span. The acquisition and tracking thresholds presented in this paper show that algorithms adapted from typical GNSS algorithms can match the specifications of GEO missions. Therefore, there is no need to implement orbital navigation filter that can interpolate the receiver position. These filters imply the development of complex algorithms based on a highly precise state model and which take into account error model (solar pressure, clock bias…) in order to ensure the performances.

A model of the receiver is under development using new hardware-software codesign methodologies. This receiver is reconfigurable in order to accept the current signals as well as future ones. The on-going step of the project is the implementation of the receiver on a FPGA prototyping board. Several implementation solutions will be compared in term of power consumption, latency and area.